\documentclass{article}

\usepackage{PRIMEarxiv}

\usepackage[utf8]{inputenc} 
\usepackage[T1]{fontenc}    
\usepackage{hyperref}       
\usepackage{url}            
\usepackage{booktabs}       
\usepackage{amsfonts}       
\usepackage{nicefrac}       
\usepackage{microtype}      
\usepackage{lipsum}
\usepackage{fancyhdr}       
\usepackage{graphicx}  
\usepackage{amsmath}    
\graphicspath{{media/}}     
\usepackage{enumerate}

\title{Transfer learning method in the problem of binary classification of chest X-rays
}

\author{
  Kolesnikov Dmitry, \\
   Bauman Moscow State Technical University \\
  Russia, Moscow\\
  \texttt{koldim2001@gmail.com} 
}

\begin{document}
\maketitle

\begin{abstract}

The possibility of high-precision and rapid detection of pathologies on chest X-rays makes it possible to detect the development of pneumonia at an early stage and begin immediate treatment. Artificial intelligence can speed up and qualitatively improve the procedure of X-ray analysis and give recommendations to the doctor for additional consideration of suspicious images. The purpose of this study is to determine the best models and implementations of the transfer learning  method in the binary classification problem in the presence of a small amount of training data. In this article, various methods of augmentation of the initial data and approaches to training ResNet and DenseNet models for black-and-white X-ray images are considered, those approaches that contribute to obtaining the highest results of the accuracy of determining cases of pneumonia and norm at the testing stage are identified.
\end{abstract}


\section{Introduction}
Pneumonia is one of the most common infectious diseases that affects millions of people worldwide, causing complications to the lungs and leading to significant morbidity and mortality \cite{kulkarni2016pneumonia}. Chest X-ray imaging is a crucial tool for diagnosing pneumonia, which can be a complex and time-consuming process. The current COVID-19 pandemic has further increased the workload of hospitals, making it even more challenging for doctors to accurately diagnose and treat pneumonia in a timely and efficient manner \cite{dinh2021overcrowding}.

Interpreting chest X-ray images requires specialized medical training and experience, and the quality of the images can significantly affect the accuracy of the diagnosis \cite{jaiswal2019identifying}. The interpretation of chest X-rays is subjective and can lead to significant variability among radiologists, which can result in missed or delayed diagnoses \cite{pezzotti2014chest}. Additionally, due to the high volume of patients with pneumonia and other lung infections, doctors may not have enough time to review all X-ray images in detail, leading to potential errors and delays in diagnosis.

To address these challenges, there is a growing interest in developing automated approaches to assist medical professionals in the interpretation of chest X-ray images . Such systems can reduce the time and workload for radiologists, improve the accuracy and consistency of diagnoses, and ultimately lead to better patient outcomes \cite{narin2021automatic}, \cite{ke2019neuro}.

Recent studies have shown promising results in developing automated approaches to classify chest X-ray images for pneumonia \cite{2021classification}, \cite{li2020accuracy}. For example, deep learning algorithms have been used to analyze X-ray images and differentiate between normal and abnormal cases, with high accuracy \cite{kundu2021pneumonia}, \cite{sarki2022automated}. Other studies have focused on detecting specific features associated with pneumonia, such as consolidation or infiltrates, to improve diagnostic accuracy \cite{punn2021automated}, \cite{punn2021automated}, \cite{soares2023evaluation}.

Despite the promising results in automated approaches for pneumonia diagnosis using chest X-ray images, there are challenges that need to be addressed. One major challenge is the variability in image format and quality among different medical organizations, making it difficult to develop a universal approach for image analysis and classification.

To address this challenge, researchers are exploring approaches that can be tailored to the specific features of chest X-ray images in a given medical organization \cite{ajmi2020using}. This requires the development of a system that can be trained on a limited number of images from that organization, while still achieving high levels of accuracy in diagnosis \cite{apostolopoulos2020covid}.

In conclusion, the need for accurate and efficient diagnosis of pneumonia is critical, especially during the current pandemic and other seasonal infectious disease outbreaks. Developing automated approaches to support medical decision-making in this area is essential to improve the quality and efficiency of care. Tailoring these approaches to the specific features of images in a given medical organization is essential for achieving high levels of accuracy in diagnosis. Further research in this field is necessary to optimize these systems and ensure their effectiveness in real-world clinical settings.

\section{Related Work}
\label{sec:headings}
In recent years, there has been an increasing interest in the use of artificial intelligence (AI) techniques for the automatic classification of chest X-ray images. This is due to the potential for AI to provide accurate and efficient diagnoses, especially in situations where there is a shortage of radiologists or a large number of images to be analyzed.

Several approaches have been proposed for the classification of chest X-ray images, including traditional machine learning algorithms such as support vector machines (SVM) and deep learning techniques such as convolutional neural networks (CNNs). While traditional machine learning algorithms have shown some success, deep learning techniques have emerged as the state-of-the-art approach for this task \cite{mahdy2020automatic}, \cite{apostolopoulos2020covid}.

However, one of the challenges of deep learning techniques is the need for large amounts of data to train a model. In the case of chest X-ray images, obtaining a large dataset can be difficult due to privacy concerns and the difficulty of obtaining labeled data. Therefore, transfer learning has emerged as the best approach for classifying images in the presence of a small number of training photos \cite{apostolopoulos2020covid}.

Transfer learning is a deep learning technique that involves using a pre-trained model, which has already been trained on a large dataset, and fine-tuning it for a new task with a smaller dataset. The pre-trained model is usually trained on a large dataset, such as ImageNet, which contains millions of labeled images. The model is then adapted to the new task by training it on a smaller dataset, such as a dataset of chest X-ray images, while keeping the earlier learned weights fixed \cite{zhuang2020comprehensive}.

The advantages of transfer learning include the ability to use a pre-trained model, which has already learned important features from a large dataset, and the ability to train a model with a smaller dataset. This approach can also help to reduce the risk of overfitting, which can occur when training a model with a small dataset \cite{mahajan2019towards}.

However, the disadvantages of transfer learning include the possibility of the pre-trained model not being suitable for the new task, and the need for a large pre-trained model, which can be computationally expensive to train. Additionally, fine-tuning a pre-trained model requires careful selection of hyperparameters and can be a time-consuming process \cite{zhuang2020comprehensive}.

There has been a significant amount of research conducted in the field of transfer learning for classifying the presence of pneumonia in chest X-ray images. Deep learning specialists have explored various transfer learning techniques and achieved high levels of accuracy in the classification task \cite{narin2021automatic}, \cite{manickam2021automated}.

However, the challenge of working with a small number of source images still remains. To address this challenge, a unique approach was proposed in this study that involved a special approach to image preprocessing and augmentation, as well as the implementation of multiple approaches that differ in their use of pre-trained networks with three input channels, even when the dataset provided black-and-white photos.

The proposed approach builds upon the existing research in the field of transfer learning by incorporating novel techniques to overcome the limitations of a small dataset. Specifically, the study focuses on image preprocessing and augmentation, which are critical steps in the training process for deep learning models. By utilizing these techniques, the proposed approach aims to improve the accuracy of the classification task and mitigate the risk of overfitting.

The implementation of multiple approaches that differ in their use of pre-trained networks with three input channels is another key contribution of this study. This allows for a comparison of the effectiveness of different pre-trained networks and provides insights into which networks are best suited for this task. 

This will allow applied specialists to immediately train models for the dataset of a particular hospital, bypassing the stage of long-term research and selection of architectures of the transfer learning approach.

\section{Methods}
To conduct the study, a publicly available (open source) dataset of X-ray images "Chest X-Ray Images Pneumonia" was taken from the Kaggle website. This dataset contains the train images section, which has 1341 images of norm and 3875 images with pneumonia.

\subsection{Image pre-processing and augmentation}
More than 95\% of the images in the dataset are black and white, but there are still, albeit rarely, three-channel RGB images. Therefore, the first stage of processing was the translation of all images into a single-channel format with averaging of channel intensities for color images.

In the validation dataset, only 6 images of each class were initially presented, which is disproportionately too small to assess the quality of the network when selecting hyperparameters. So the test and validation datasets were concatenated and then evenly divided in half. Thus, in the test and validation subsample, the number of images was 320 in each. At the same time, the proportion between classes is approximately the following: 60\% class pneumonia and 40\% class norm.

Since the number of images with pathology in the training dataset significantly exceeded the number of images without it, it was necessary to reduce such a disparity for better training. Since the initial goal of the work was to study the best principles of network training for classifying chest X-ray images on a small number of training images, therefore, the dowsamling approach of objects from the pneumonia class was chosen. After the implementation of this method, the resulting dataset began to contain 1341 images of the norm and the same number of pathology.

After that, the values of the mean and standard deviation of intensity in the training sample were determined. These values will be used when implementing input data normalization during network training and testing.

In order to increase the size of the resulting training dataset of images, the augmentation method was employed. This technique involves generating new examples of data based on the existing ones by applying various transformations \cite{wang2017effectiveness}. Specifically, four different types of transformations were implemented in this study to augment the dataset.

\begin{figure}
  \centering
  \includegraphics[width=12cm]{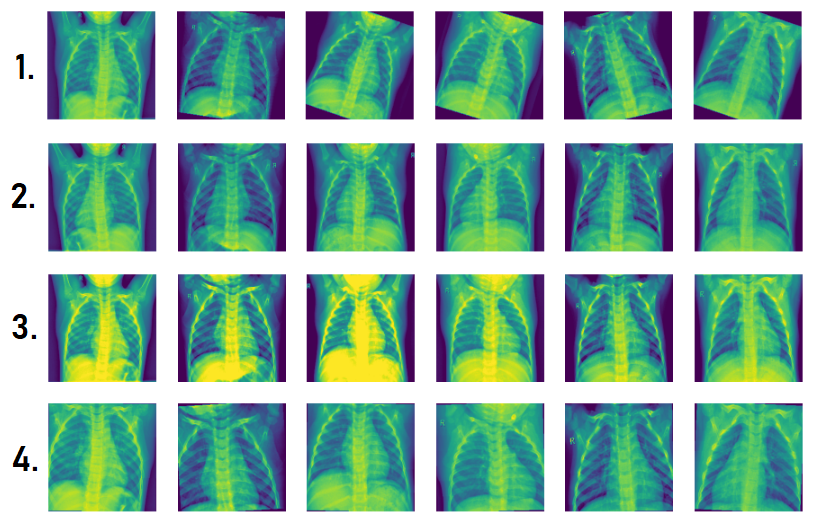}
  \caption{Transformations for augmentation.}
  \label{fig:fig_tr}
\end{figure}

These types of transformations that have been implemented:
\begin{enumerate}
\item Changing the rotation angle of the image to a random angle from 0 to 20 degrees. Next, the image is compressed to a size of 224 by 224 pixels (to match the input sizes of ResNet and DenseNet networks). And at the last stage normalization of the resulting images.

\item Symmetrical mapping around the vertical. Next, the image is compressed to a size of 224 by 224 pixels (to match the input sizes of ResNet and DenseNet networks). And at the last stage normalization of the resulting images.

\item Change the brightness, contrast and saturation parameters. Next, the image is compressed to a size of 224 by 224 pixels (to match the input sizes of ResNet and DenseNet networks). And at the last stage normalization of the resulting images.

\item First, the image is compressed to the size of 280 by 280, then a part of the size 224 by 224 is randomly cut out, then with a probability of 0.5 there will be a reflection around the horizontal as in transformation No. 2 and then a small random rotation from 0 to 5 degrees. And at the very end, normalization is implemented
\end{enumerate}

Examples of the results of the described transformation are presented in Figure \ref{fig:fig_tr}.

Different types of transformations have been implemented since the augmentation method has been shown to be an effective way of increasing the diversity of the dataset and improving the performance of machine learning models. However, it is important to note that using too many or inappropriate types of transformations can lead to overfitting and reduced performance. Therefore, the types and number of transformations used in this study were carefully selected to strike a balance between increasing the dataset size and maintaining the model's ability to generalize.

A study was conducted in which a combination of initial images on a training dataset was implemented together with each type of transformation separately. The highest quality indicators on the test dataset after training a custom network were shown by a network trained on a dataset consisting of initial training images and photos symmetrically reflected relative to the vertical (transformation №2). Therefore, it was decided to use this type of augmentation, which will increase the number of images in training by 2 times.

In total, after the preprocessing and augmentation stage, a training dataset was obtained, consisting of 5364 X-ray images, half of which are normal and the second half are pictures with pneumonia.

\subsection{Implementation of transfer learning and selection of hyperparameters and model's training structure }
ResNet18 and DenseNet121 are both popular deep neural network architectures used in computer vision tasks such as image classification, object detection, and segmentation.

ResNet18 is a variant of the ResNet (Residual Network) architecture, introduced by Microsoft Research in 2015. ResNet18 consists of 18 layers of convolutional neural network (CNN) and has skip connections between layers, which allow the network to be trained more efficiently and to reduce the vanishing gradient problem. ResNet18 has been widely used in transfer learning, where the pre-trained model is fine-tuned on a new dataset with a small number of training images. One of the advantages of ResNet18 is that it can achieve high accuracy with a relatively small number of parameters, making it suitable for deployment on devices with limited computational resources \cite{al2021automatic}.

DenseNet121, on the other hand, is a more recent architecture introduced by the University of Waterloo in 2016. The name "DenseNet" comes from the fact that the network is densely connected, meaning that each layer receives input not only from its previous layer but also from all previous layers in the network. This allows the network to learn more complex representations and achieve better performance with fewer parameters. DenseNet121 consists of 121 layers of CNN and has been shown to outperform other state-of-the-art architectures on various computer vision tasks \cite{ezzat2020gsa}.

Both of these convolutional neural network architectures were considered in this study on the classification of chest X-Ray Images.

The images that were used to train the networks were black and white, so it became necessary to come up with a concept for working with them, since the original architectures of the DenseNet and ResNet networks imply three-channel images with a size of 224 by 224 pixels at the input.

So 2 different approaches to working with source images were considered:

\begin{itemize}

\item Leave the images single-channel and redefine the first convolutional layer, making the value of the number of input channels equal to 1.
\item Make the pictures three-channel by copying the value of one original black-and-white channel. That is, the artificial creation of two other channels.
\end{itemize}

To determine the best network training strategy, 3 experiments were conducted separately for the ResNet and DenseNet networks. At the same time, the pre-trained weights of neurons were initially uploaded to the network.

\begin{enumerate}[I]

    \item Fixed all convolutional layers completely and only trained fully connected layers.\\ \\ At the same time, I changed the architecture of the fully connected $fc_{\ new}$ layer. I made one hidden layer with the number of neurons, which will be used as a matching hyperparameter. The best hyperparameter was determined by $n_{\ neurons}$ based on accuracy results on the validation sample.\\ Since the photos are the original black and white, when creating training datasets, I duplicated the channel values so that the number of input channels of the convolutional network was 3 as in the original network (3, 224, 224).
    
    \item Fixed all convolutional layers except the first input, in which we set the number of input channels to 1 instead of 3 and trained fully connected layers.\\ \\ At the same time, I changed the architecture of the fully connected $fc_{\ new}$ layer. I made one hidden layer with the number of neurons, which will be used as a matching hyperparameter. The best hyperparameter was determined by $n_{\ neurons}$ based on accuracy results on the validation sample.\\ Since I changed the number of channels at the network input, so I trained the network on ordinary black and white images, the size of (1, 224, 224).
    
    \item Fully trained network without changing anything in it (except for the output number of neurons in the fully connected part).\\ \\ Since the photos are the original black and white, when creating training datasets, I duplicated the channel values so that the number of input channels of the convolutional network was 3 as in the original network (3, 224, 224).
\end{enumerate}

\begin{figure}
  \centering
  \includegraphics[width=7cm]{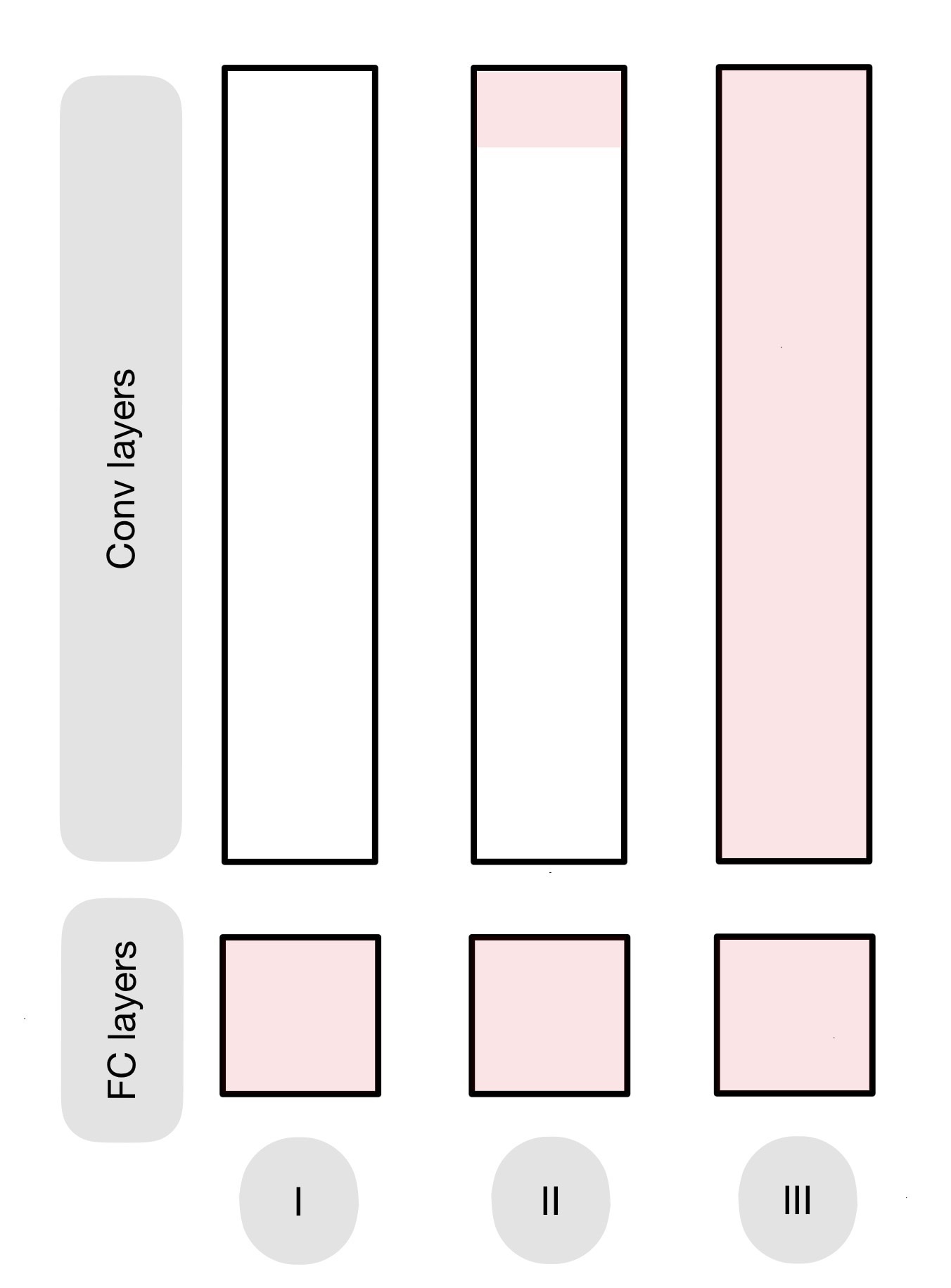}
  \caption{The transfer learning training strategies.}
  \label{fig:fig_l}
\end{figure}
The layers available for training, that is, in which weights can change due to gradient descents, are represented in pink on Figure \ref{fig:fig_l}.

\vspace{15mm} 

In experiments I and II, fully connected layers were modified with this architecture:
\begin{alignat*}{1}
fc_{\ new} =  Sequential(&Dropout(p=0.2),  \\
         &    Linear(n_{\ input}, \  n_{\ neurons}),\\
         &    ReLU(),\\
         &    Linear(n_{\ neurons}, \ 2))
\end{alignat*}

 Thus, there was an intermediate layer, which had a number of neurons taking the values of 10, 100 and 500. In the process of conducting these experiments for each of the ResNet-18 and DenseNet-121 networks, I selected the best given hyperparameter.

I also conducted an additional experiment in which I trained networks without using transfer learning, so I didn't use pre-trained weights. I thereby wanted to find out what would change with full training without pre-selected knowledge.

 The size of the mini-batch was chosen to be equal to 30 images, so one epoch of training with the number of images equal to 5364 was equal to performing 179 gradient descents. During the training, the learning rate scheduler was used, which after every 5 epochs of training reduced the learning rate by 10 times, thereby increasing the chances of falling into the global minimum of the loss function, which was cross entropy.

\section{Results}

The solution of the project described in this article is presented on the open source platform GitHub in the public repository at the link:
\begin{center}
  \url{https://github.com/Koldim2001/transfer_learning_CNN}
\end{center}

As a result, 3 networks (one for each experiment) and 3 networks using DenseNet-121 were implemented and trained using the transfer learning method for ResNet-18. For the I and II experiments, the dependences of the accuracy metric change on validation and training for each of the three hyperparameter values were obtained.

\begin{figure}
  \centering
  \includegraphics[width=17cm]{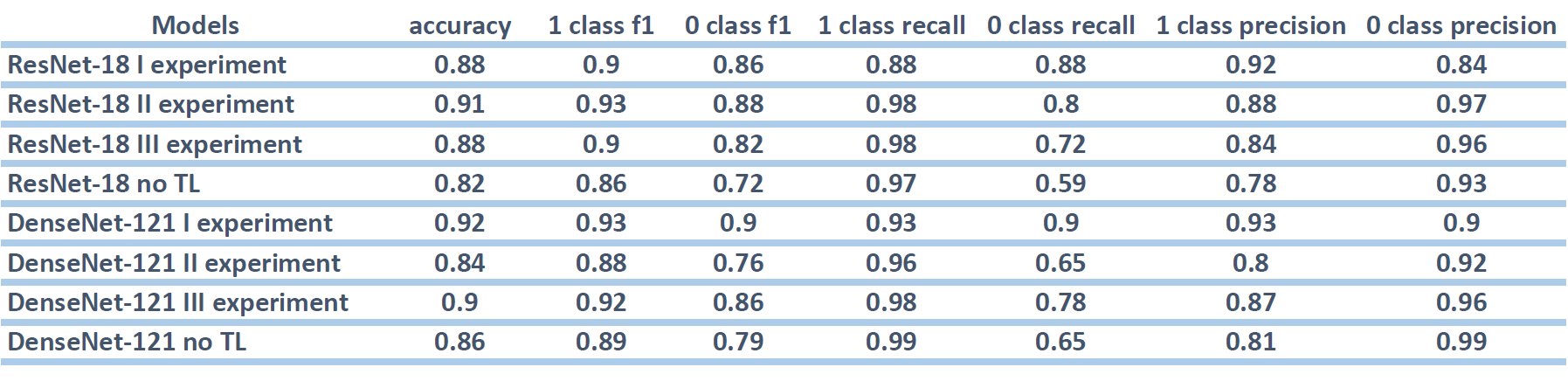}
  \caption{Metrics of different models.}
  \label{fig:fig_tab}
\end{figure}

As a result, for each experiment, the value of the hyperparameter and the number of the training epoch were found, on which it was possible to obtain the highest accuracy value on validation. As a result, a common table was formed for the best networks of each experiment with the resulting metric values for each class (1 – pneumonia, 0 – norm). The results can be seen in Figure \ref{fig:fig_tab}.

Also, the results of ResNet-18 and DenseNet-121 network training have been added to the table with full training without pre-loaded weights, that is, without the use of transfer learning (no TL).

Graphs of the dependence of changes in the accuracy metric for validation and training on the number of training epochs were also obtained. In this way, 2 graphs were obtained for each of the ResNet-18 and DenseNet-121 networks. These graphs are presented in Figures \ref{fig:res_train},\ref{fig:res_val},\ref{fig:dense_train} and \ref{fig:dense_val}. And will be analyzed in detail in the section of the article “Discussion".

\section{Discussion}
\subsection{ResNet-18}

\begin{figure}
  \centering
  \includegraphics[width=12.5cm]{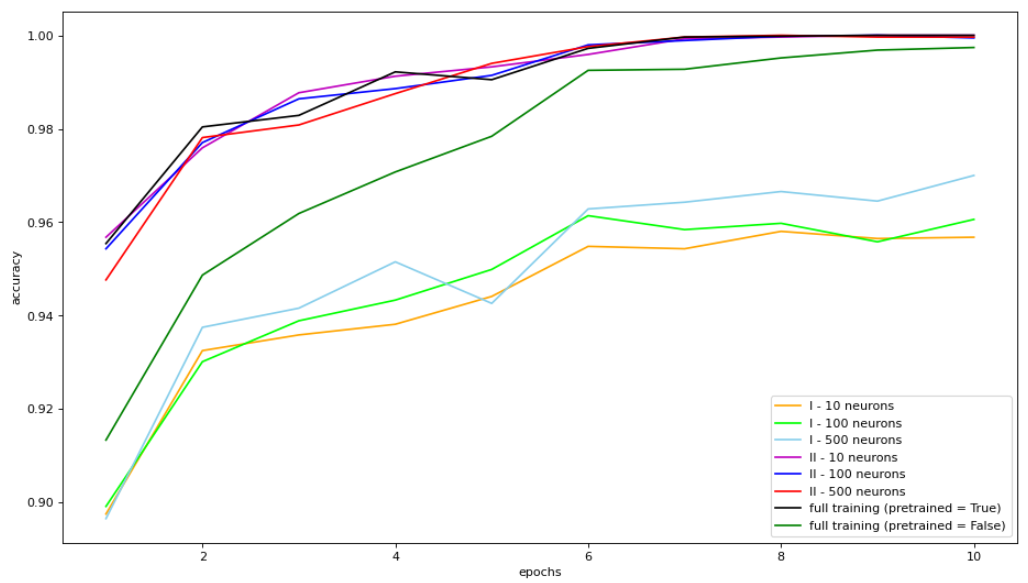}
  \caption{Accuracy on the training dataset for various types of ResNet training.}
  \label{fig:res_train}
\end{figure}

\begin{figure}
  \centering
  \includegraphics[width=12.5cm]{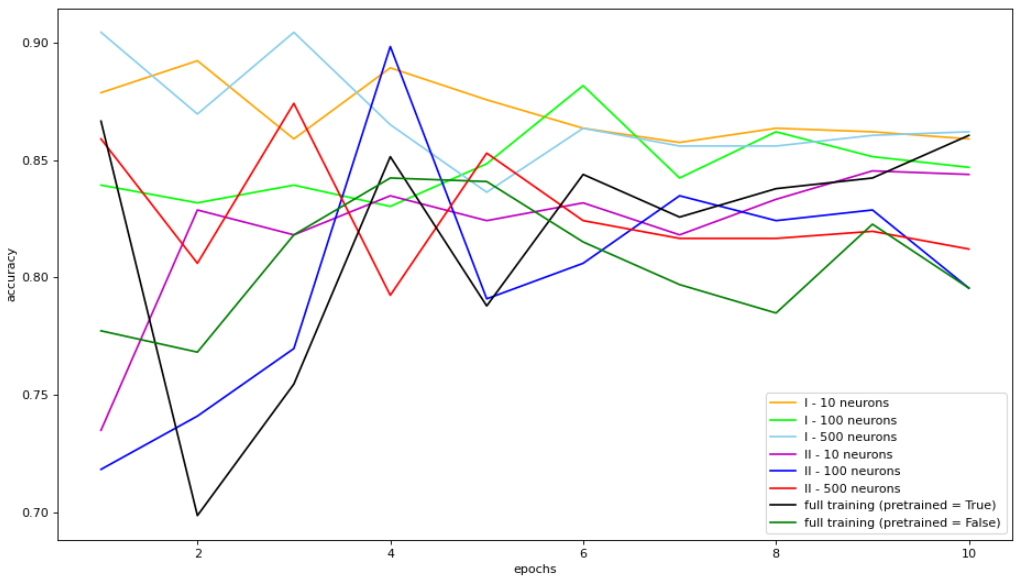}
  \caption{Validation accuracy in various types of ResNet training.}
  \label{fig:res_val}
\end{figure}

As a result of the study, it was revealed that the best accuracy results on the test were achieved when using a model with a modified first convolutional layer (II experiment) and a fully connected part with the number of neurons of the intermediate layer equal to 100 (90\% accuracy). 

According to the Figure \ref{fig:res_val}, in the first experiment (I), for any values of the hyperparameter, $n_{ neurons}$ behaves more stably during validation and has low variability relative to the trend during training (while the accuracy values themselves are consistently higher than in all other experiments). For this reason, it is possible say that the principle of training with the fixation of all convolutional layers has shown better results than all the other considered approaches to ResNet training for this binary classification.

So such a high result on the test in the second experiment may be caused, among other things, by a coincidence (the test sample is too small to consider the difference of 2\% on accuracy statistically significant).

It was also found that the use of pre-trained weights of neurons significantly accelerates the learning rate. In additional experiment, I trained the ResNet18 network without pre-selected weights and received longer training on the train dataset (Figure \ref{fig:res_train}) as well as lower accuracy rates on validation (Figure \ref{fig:res_val}). This confirms the usefulness of using transfer learning technology - the application of knowledge gained in solving one problem to solve a similar one.

In addition, it was noted that at the 6th epoch of training, the quality of validation for most models reached a plateau and stopped jumping due to random coincidences of answers. However, with an increase in the number of neurons on the hidden layer of a fully connected network, the speed of overfitting increases. So the optimal value of learning epochs directly depends on this hyperparameter.

When creating a medical decision support system, an important parameter in assessing the quality of the model is f1 score and recall. It is important for us not to miss patients with pneumonia and to accurately name all patients as sick. For the best models of each of the experiments, the values f1, recall and precision for class 1 (presence of pneumonia) were determined.\\
When analyzing the results of the experiment, it can be seen that the model from the second experiment has the highest f1 score = 0.93 and recall = 0.98, while the other 2 transfer learning models have f1 score = 0.90 (recall of the first experiment 0.88, the second 0.98). Since the best model for the experiment was chosen based on the result of the highest accuracy during validation during training, and the validation sample was dominated by Class 1 photos, therefore, high recall values are predictable, so it would be correct to evaluate it by f1 score.

When training without pre-trained network weights, we obtained the lowest values of the f1 metric (0.86), which once again shows the advantage of transfer learning in this classification problem with a small train dataset size.

\subsection{DenseNet-121}

As a result of the study, it was revealed that the best accuracy results on the test were achieved when using a pre-trained model with only a fully connected layer modified (experiment I) with the number of intermediate layers equal to 10 (92\% accuracy). 

\begin{figure}
  \centering
  \includegraphics[width=12.5cm]{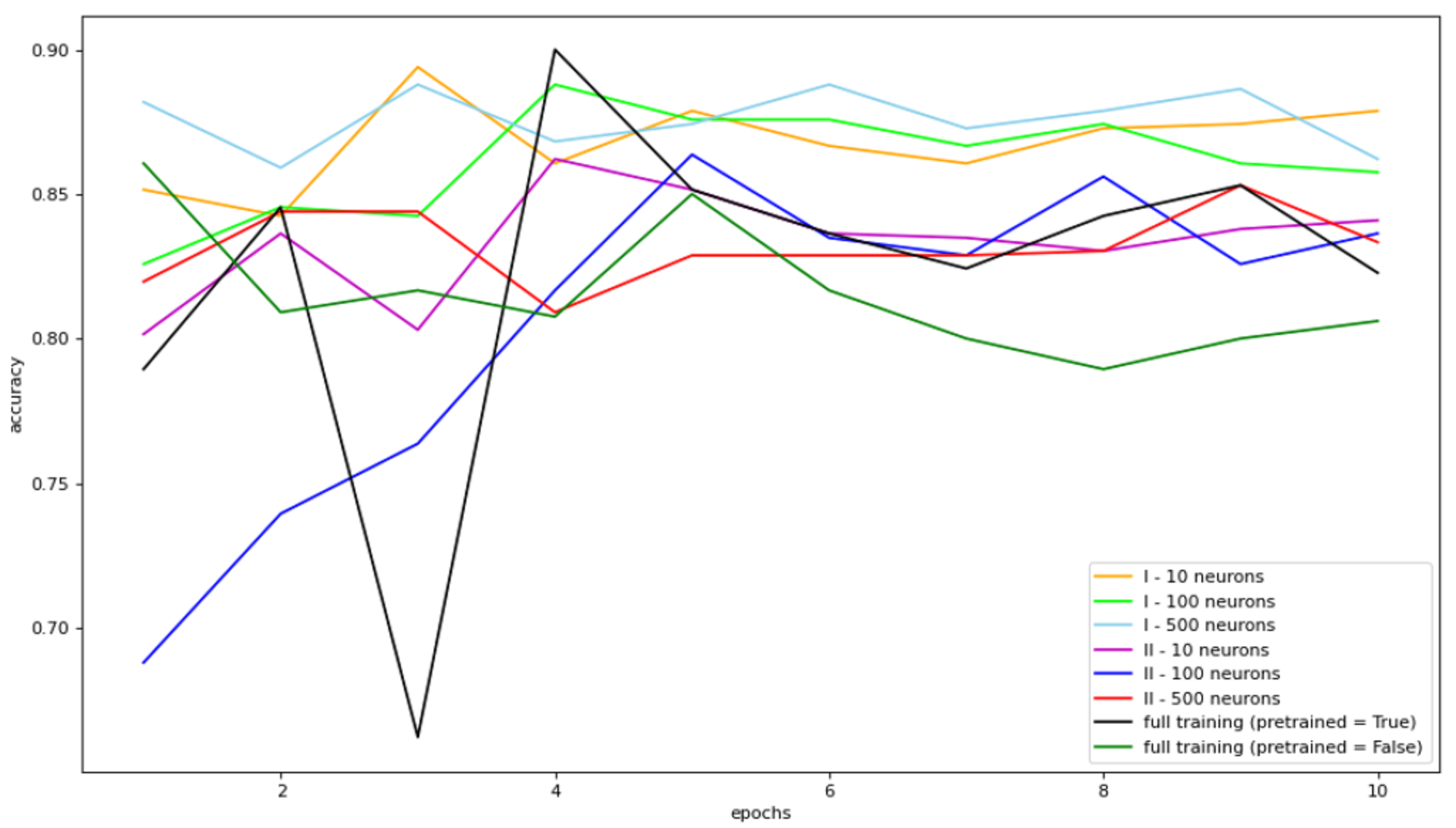}
  \caption{Validation accuracy in various types of DenseNet training.}
  \label{fig:dense_val}
\end{figure}

According to the Figure \ref{fig:dense_val}, the first experiment at any values of the $n_{ neurons}$ hyperparameter behaves more stably during validation and has low variability relative to the trend during training (while the accuracy values themselves are consistently higher than those of all other experiments). For this reason, we can say that the principle of training with the fixation of all convolutional layers showed better results than all the other DenseNet training approaches considered for this binary classification.

\begin{figure}
  \centering
  \includegraphics[width=12.5cm]{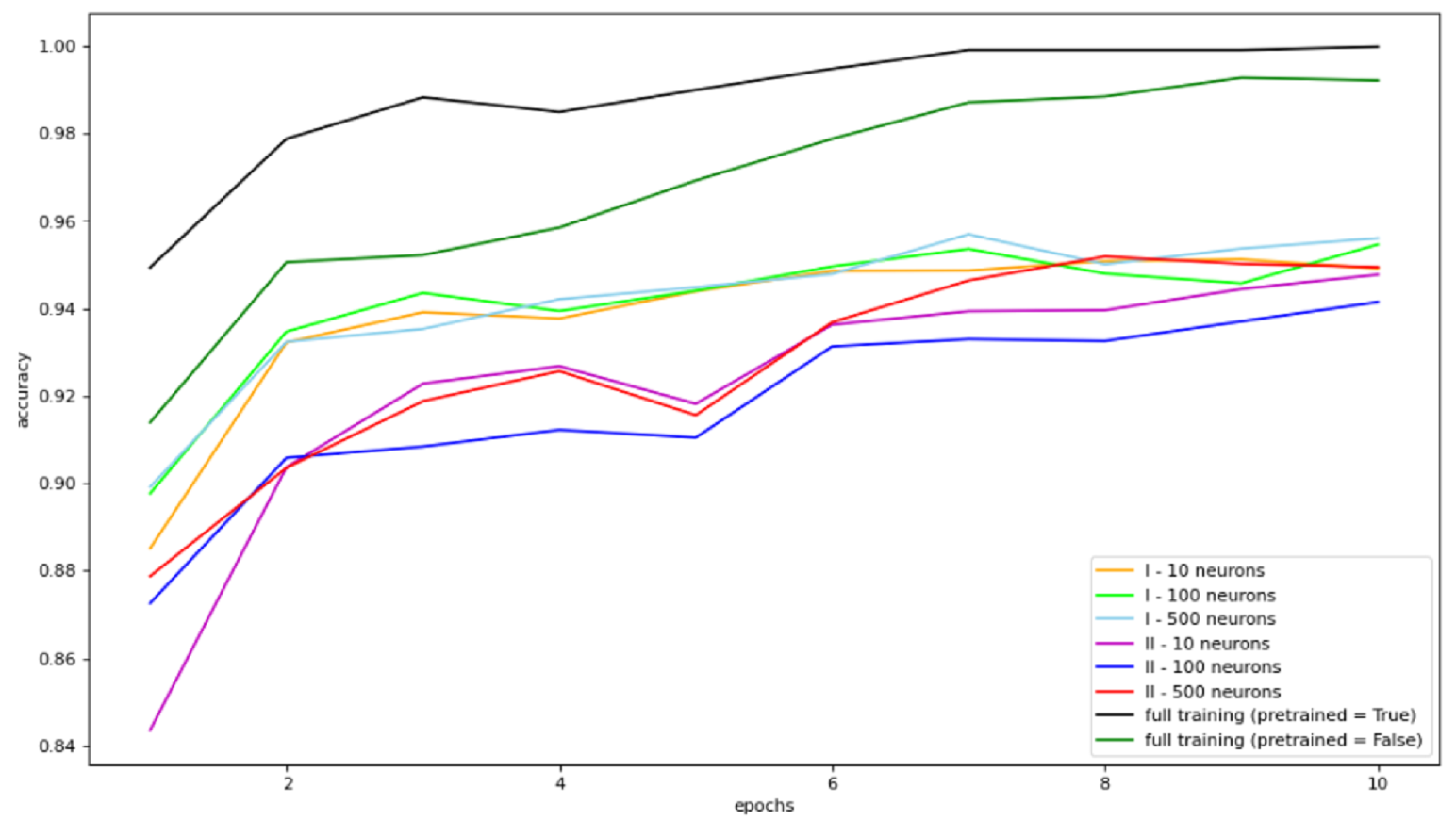}
  \caption{Accuracy on the training dataset for various types of DenseNet training.}
  \label{fig:dense_train}
\end{figure}

It was also found that the learning speed in the second experiment (II) at any hyperparameter values is significantly lower than in the first and 3rd experiments (Figure \ref{fig:dense_train}). So a network of 2 experiments needs to be trained for at least 6 epochs in order to achieve consistently high results on validation. Up to 6 epochs, the accuracy value randomly jumps from 70 to 90, which does not say anything good about the quality of the model.

In addition, it was noted that with full training, there is a very rapid overfitting, which is why the accuracy value on the train becomes approximately 100\% at the 6th epoch, while the results on validation are only beginning to decrease. For this reason, training the entire network is not the right solution.

As it was said earlier, when creating a medical decision support system, an important parameter in assessing the quality of the model is f1 score and recall for pneumonia class. When analyzing the results of the experiment, it can be seen that the model from the first experiment has the highest f1 score = 0.93, but at the same time recall is only 0.93, while all other models have f1 no higher than 0.9, but higher recall class pneumonia values. Since the best model for the experiment was chosen based on the result of the highest accuracy during validation during training, and the validation samples was dominated by Class 1 photos, therefore, high recall values are predictable, so it would be correct to evaluate it by f1 score.

\section{Conclusion}
Summing up, the application of the transfer learning method when training convolutional neural networks in the chest X-ray image classification task showed high accuracy values on the test. 

As a result of verified research, it can be said that the DenseNet network showed the highest results compared to ResNet. Namely, when learning with a fixed convolutional part. And in order for black-and-white images to be run through such a network, it is first necessary to duplicate the values of a single channel in order to make the number of channels of the input layer equal to three (as the pre-trained network needs). 

At the same time, for any values of hyperparameters (the number of neurons of the intermediate layer), fairly high results are obtained for such a network with an accuracy of more than 87\%. In the current experiment conducted with this set of training data, the favorite was a network with the number of neurons in a fully connected layer equal to 10. This is due to the fact that such a network had a small number of selected weights and was less susceptible to overfitting Such a network learns fairly quickly and does not require more than 5 training epochs (179 gradient descents per epoch) to obtain consistently high test results.

\bibliographystyle{unsrt}  
\bibliography{references}

\begin{thebibliography}{10}

\bibitem{kulkarni2016pneumonia}
Hrishikesh Kulkarni.
\newblock What is pneumonia?
\newblock {\em American Journal of Respiratory and Critical Care Medicine},
  193(1):I, 2016.

\bibitem{dinh2021overcrowding}
Michael~M Dinh and Saartje Berendsen~Russell.
\newblock Overcrowding kills: How covid-19 could reshape emergency department
  patient flow in the new normal.
\newblock {\em Emergency Medicine Australasia}, 33(1):175--177, 2021.

\bibitem{jaiswal2019identifying}
Amit~Kumar Jaiswal, Prayag Tiwari, Sachin Kumar, Deepak Gupta, Ashish Khanna,
  and Joel~JPC Rodrigues.
\newblock Identifying pneumonia in chest x-rays: A deep learning approach.
\newblock {\em Measurement}, 145:511--518, 2019.

\bibitem{pezzotti2014chest}
William Pezzotti.
\newblock Chest x-ray interpretation: not just black and white.
\newblock {\em Nursing2022}, 44(1):40--47, 2014.

\bibitem{narin2021automatic}
Ali Narin, Ceren Kaya, and Ziynet Pamuk.
\newblock Automatic detection of coronavirus disease (covid-19) using x-ray
  images and deep convolutional neural networks.
\newblock {\em Pattern Analysis and Applications}, 24:1207--1220, 2021.

\bibitem{ke2019neuro}
Qiao Ke, Jiangshe Zhang, Wei Wei, Dawid Po{\l}ap, Marcin Wo{\'z}niak, Leon
  Ko{\'s}mider, and Robertas Dama{\v{s}}ev{\u\i}cius.
\newblock A neuro-heuristic approach for recognition of lung diseases from
  x-ray images.
\newblock {\em Expert systems with applications}, 126:218--232, 2019.

\bibitem{2021classification}
Asmaa Abbas, Mohammed~M Abdelsamea, and Mohamed~Medhat Gaber.
\newblock Classification of covid-19 in chest x-ray images using detrac deep
  convolutional neural network.
\newblock {\em Applied Intelligence}, 51:854--864, 2021.

\bibitem{li2020accuracy}
Yuanyuan Li, Zhenyan Zhang, Cong Dai, Qiang Dong, and Samireh Badrigilan.
\newblock Accuracy of deep learning for automated detection of pneumonia using
  chest x-ray images: A systematic review and meta-analysis.
\newblock {\em Computers in Biology and Medicine}, 123:103898, 2020.

\bibitem{kundu2021pneumonia}
Rohit Kundu, Ritacheta Das, Zong~Woo Geem, Gi-Tae Han, and Ram Sarkar.
\newblock Pneumonia detection in chest x-ray images using an ensemble of deep
  learning models.
\newblock {\em PloS one}, 16(9):e0256630, 2021.

\bibitem{sarki2022automated}
Rubina Sarki, Khandakar Ahmed, Hua Wang, Yanchun Zhang, and Kate Wang.
\newblock Automated detection of covid-19 through convolutional neural network
  using chest x-ray images.
\newblock {\em Plos one}, 17(1):e0262052, 2022.

\bibitem{punn2021automated}
Narinder~Singh Punn and Sonali Agarwal.
\newblock Automated diagnosis of covid-19 with limited posteroanterior chest
  x-ray images using fine-tuned deep neural networks.
\newblock {\em Applied Intelligence}, 51(5):2689--2702, 2021.

\bibitem{soares2023evaluation}
Thiego~Ramon Soares, Roberto~Dias de~Oliveira, Yiran~E Liu, Andrea
  da~Silva~Santos, Paulo Cesar~Pereira dos Santos, Luma Ravena~Soares Monte,
  Lissandra~Maia de~Oliveira, Chang~Min Park, Eui~Jin Hwang, Jason~R Andrews,
  et~al.
\newblock Evaluation of chest x-ray with automated interpretation algorithms
  for mass tuberculosis screening in prisons: A cross-sectional study.
\newblock {\em The Lancet Regional Health-Americas}, 17:100388, 2023.

\bibitem{ajmi2020using}
Chiraz Ajmi, Juan Zapata, Jos{\'e}~Javier Mart{\'\i}nez-{\'A}lvarez, Gin{\'e}s
  Dom{\'e}nech, and Ram{\'o}n Ruiz.
\newblock Using deep learning for defect classification on a small weld x-ray
  image dataset.
\newblock {\em Journal of Nondestructive Evaluation}, 39:1--13, 2020.

\bibitem{apostolopoulos2020covid}
Ioannis~D Apostolopoulos and Tzani~A Mpesiana.
\newblock Covid-19: automatic detection from x-ray images utilizing transfer
  learning with convolutional neural networks.
\newblock {\em Physical and engineering sciences in medicine}, 43:635--640,
  2020.

\bibitem{mahdy2020automatic}
Lamia~Nabil Mahdy, Kadry~Ali Ezzat, Haytham~H Elmousalami, Hassan~Aboul Ella,
  and Aboul~Ella Hassanien.
\newblock Automatic x-ray covid-19 lung image classification system based on
  multi-level thresholding and support vector machine.
\newblock {\em MedRxiv}, pages 2020--03, 2020.

\bibitem{zhuang2020comprehensive}
Fuzhen Zhuang, Zhiyuan Qi, Keyu Duan, Dongbo Xi, Yongchun Zhu, Hengshu Zhu, Hui
  Xiong, and Qing He.
\newblock A comprehensive survey on transfer learning.
\newblock {\em Proceedings of the IEEE}, 109(1):43--76, 2020.

\bibitem{mahajan2019towards}
Sarang Mahajan, Urmil Shah, Rucha Tambe, Mohit Agrawal, and Bhushan Garware.
\newblock Towards evaluating performance of domain specific transfer learning
  for pneumonia detection from x-ray images.
\newblock In {\em 2019 IEEE 5th international conference for convergence in
  technology (I2CT)}, pages 1--6. IEEE, 2019.

\bibitem{manickam2021automated}
Adhiyaman Manickam, Jianmin Jiang, Yu~Zhou, Abhinav Sagar, Rajkumar
  Soundrapandiyan, and R~Dinesh~Jackson Samuel.
\newblock Automated pneumonia detection on chest x-ray images: A deep learning
  approach with different optimizers and transfer learning architectures.
\newblock {\em Measurement}, 184:109953, 2021.

\bibitem{wang2017effectiveness}
Jason Wang, Luis Perez, et~al.
\newblock The effectiveness of data augmentation in image classification using
  deep learning.
\newblock {\em Convolutional Neural Networks Vis. Recognit}, 11(2017):1--8,
  2017.

\bibitem{al2021automatic}
Ruaa~A Al-Falluji, Zainab~Dalaf Katheeth, and Bashar Alathari.
\newblock Automatic detection of covid-19 using chest x-ray images and modified
  resnet18-based convolution neural networks.
\newblock {\em Computers, Materials, \& Continua}, pages 1301--1313, 2021.

\bibitem{ezzat2020gsa}
Dalia Ezzat, Hassan~Aboul Ella, et~al.
\newblock Gsa-densenet121-covid-19: a hybrid deep learning architecture for the
  diagnosis of covid-19 disease based on gravitational search optimization
  algorithm.
\newblock {\em arXiv preprint arXiv:2004.05084}, 2020.

\end{thebibliography}

\end{document}